\documentclass [a4paper]{JHEP3}
\usepackage{epsfig}
\newtheorem{theorem}{Theorem}
\title{
\protect\vspace{5mm}
Fermions in the vortex background on a sphere
}
\author{J.-M.~Fr\`ere$^{1}$, M.V.~Libanov$^{1,2}$, E.Ya.~Nugaev$^{2}$, and
S.V~Troitsky$^{2}$\\
$^{1}$~Service de Physique Th\'{e}orique, CP 225,\\
  Universit\'{e} Libre de Bruxelles, B--1050, Brussels, Belgium;\\
$^{2}$~Institute for Nuclear Research of the Russian Academy of
Sciences,\\
60th October Anniversary Prospect 7a, Moscow 117312 Russia\\
E-mail: \email{frere@ulb.ac.be, ml@ms2.inr.ac.ru,
emin@ms2.inr.ac.ru, st@ms2.inr.ac.ru} }

\preprint{ULB-TH-03-13}
\abstract{
In 5+1 dimensions, we construct a vortex-like solution on a
two-dimensional sphere. We study fermionic zero modes in the background of
this solution and relate them to the replication of fermion families in
the Standard Model. In particular, using a compactified space removes the
need for the difficult localisation of gauge fields, while the present
procedure (rather than naive compactification on a disk) also removes
spurious fermionic modes.
}
\begin{document}
\section{Introduction.}
Coupling of fermions to background bosonic fields of nontrivial topology
may lead to chiral fermionic zero modes. In particular, for the case of
the Abrikosov-Nielsen-Olesen vortex this fact is widely known for several
decades~\cite{Jackiw:1981ee}.  The number of the zero modes coincides with
the topological number, that is, with the magnetic flux of the vortex. One
remarkable application of this feature is to models with Large Extra
Dimensions (LED) where chiral fermions of the Standard Model are described
by the zero modes of multi-dimensional fermions localized in the
(four-dimensional) core of a topological defect \cite{RuSha}. In
particular, this approach can explain the origin of the fermionic family
replication and the fermionic mass hierarchy. In
Refs.~\cite{Libanov:2000uf,Frere:2000dc,Frere:2001ug,Libanov:2002tm}, we
have constructed and studied models in which a single family of fermions
in six dimensions with vector-like couplings to the Standard Model (SM)
bosons gives rise to three generations of chiral Standard Model fermions
in four dimensions. We have also shown that in these models a power-like
hierarchy of the fermionic masses and mixings appears in a natural way.

One of the principal issues of  models with LED is the localization of the
Standard Model gauge fields. One of possible ways to avoid this problem is
to consider the transverse extra dimensional space as a compact manifold
and to allow gauge fields to propagate freely in the extra dimensions. It
is however not a straightforward task to put a vortex and fermionic zero
modes in a finite volume space. Indeed, in a flat space with a boundary,
"stray" fermionic zero modes of opposite chirality appear in addition to
the chiral zero modes localized in the core. This happens because of the
finite volume of the bulk: the modes which would have been killed by the
normalization condition in the case of infinite space, survive in  a
finite volume. Even imposing the physical boundary condition which
corresponds to the absence of the fermionic current outside the boundary
does not help since the current of the zero modes lack  radial components.
Another approach to the localization of gauge fields in the vortex
background is also possible. In particular,
in~\cite{Gravity,Randjbar-Daemi:2003qd} a vortex-like solution was
considered in the presence of gravity. It was shown that the localized
zero modes of vector field appear in these models.

In this paper we consider the vortex and the fermions taking a two
dimensional sphere as extra dimensions. Keeping in mind the application of
this system to the model with LED we will refer to the coordinates on the
sphere as to the fourth and the fifth coordinates on this $M^4\times S^2$
manifold, where $M^4$ represents our four-dimensional Minkowski space. Our
results, however, do not depend on the number of ordinary dimensions. We
will argue also that the main features of the "flat" models
~\cite{Libanov:2000uf,Frere:2000dc,Frere:2001ug,Libanov:2002tm} remain
intact in the spherical case.

\section{Vortex on a sphere.}
\label{Section/Pg2/1:sphere/Vortex on a sphere.}
In this section we
construct the Abrikosov-Nielsen-Olesen vortex solution on a $M^4\times
S^2$ manifold. The metric $g_{AB}$ of this manifold is determined by
\begin{equation}
ds^2=g_{AB} dx^Adx^B=g_{\mu \nu }dx^\mu dx^\nu -R^2(d\theta
^2+\sin^2\theta d\varphi ^2),
\label{metric}
\end{equation}
where $g_{\mu \nu }={\rm diag}(+,-,\ldots ,-)$ is the four dimensional
Minkowski metric, capital Latin indices are $A,B=0,\ldots 5$, Greek
indices are $\mu ,\nu =0,\ldots ,3$. To generate the vortex solution we
introduce an Abelian Higgs Lagrangian,
\begin{equation}
{\cal L}_V=\sqrt{-\det(g_{AB})}\left(-\frac{1}{4}F_{AB}F^{AB}+(D^{A}\Phi) ^\dag
D_{A}\Phi -\frac{\lambda }{2}(|\Phi |^2-v^2)^2\right),
\label{Eq/Pg2/1A:sphere}
\end{equation}
where $F_{AB}=\partial_A A_B-\partial_B A_A$, $D_A\Phi =(\partial
_A-ieA_A)\Phi $, $A_A$ and $\Phi$ are a $U(1)$ gauge field and a complex
scalar field, respectively.

The vortex solution which we are interested in is a static solution of the
field equations depenging only on $\theta$ and $\varphi $, the coordinates
on $S^2$. Let us introduce the standard ansatz for a vortex with winding
number one,
\begin{equation}
A_\varphi =\frac{1}{e}A(\theta) \,,\ \ \ A_\theta  =0\,,\ \ \ \Phi
=F(\theta ){\rm e}^{i\varphi }.
 \label{Eq/Pg2/1:sphere}
\end{equation}
We obtain the following set of the field equations,
 \begin{eqnarray}
 &&\frac{1}{\sin{\theta}}\partial_\theta
 \frac{1}{\sin{\theta}}\partial_\theta A
 -\frac{2R^2e^2}{\sin^2{\theta}}F^2(A-1)=0,
 \label{Eqn/Pg3/1A:sphere}
\\
 &&\frac{1}{\sin{\theta}}\partial_\theta\sin{\theta}\partial_\theta F
 -\frac{1}{\sin^2{\theta}}F(A-1)^2-\lambda R^2 F(F^2-v^2)=0.
\label{Eqn/Pg3/2:sphere}
 \end{eqnarray}

Let us discuss now the boundary conditions to the field equations. To
obtain a non-trivial soliton one should impose zero boundary conditions
for the fields $A(\theta )$ and $F(\theta )$ on the North pole of the
sphere ($\theta =0$) and non-zero boundary conditions on the South pole of
the sphere ($\theta =\pi $). However, the ansatz (\ref{Eq/Pg2/1:sphere})
with these boundary conditions is singular on the South pole due to the
$\varphi $-dependence. To avoid this difficulty we essentially repeat the
discourses by Wu and Yang~\cite{Wu:es}. We introduce two patches on the
sphere and a different ansatz in the Southern patch,
\begin{equation}
A_\varphi^S =\frac{1}{e}A^S(\theta) \,,\ \ \ A_\theta^S =0\,,\ \ \ \Phi
=F^S(\theta ),
\label{Eq/Pg3/1A:sphere}
\end{equation}
where
\begin{equation}
A^S(\pi )=0\,,\ \ \ \partial _\theta F(\pi )=0.
\label{Eq/Pg3/2:sphere}
\end{equation}

In the overlapping region the two ansatzes (\ref{Eq/Pg2/1:sphere}) and
(\ref{Eq/Pg3/1A:sphere}) are related by a gauge transformation:
\begin{equation}
F(\theta ){\rm e}^{i\varphi }={\rm e}^{i\alpha }F^S(\theta ),
\label{Eq/Pg3/3:sphere}
\end{equation}
\[
A(\theta )=A^S(\theta )+\partial _\varphi \alpha.
\]
Therefore, to satisfy the first of the equations one can choose $\alpha
=\varphi $ and, hence,
\begin{equation}
A(\theta )=A^S(\theta )+1.
\label{Eq/Pg3/5:sphere}
\end{equation}
If one introduces now another field $\Psi $ coupled to $A_A $ with the
charge $q$, thus transforming as
\[
\Psi '=\Omega \Psi =\exp\left(i\frac{q}{e}\alpha \right)\Psi,
\]
then one should require $\Omega $ to be single-valued which brings in
turn the famous Dirac's quantization condition
\begin{equation}
\frac{q}{e}=\mbox{integer}
\label{Eq/Pg4/1:sphere}
\end{equation}
This situation differs from the flat case where one might introduce fields
with the half-integer charge.

The necessity of the two patches and the appearance of the Dirac's
condition become more transparent if we examine the following relationship
between the vortex on $S^2$ and a monopole in the three dimensional space.
Let us consider a monopole with the magnetic charge $g_M$ in three
dimensions in vacuum. The field configuration of the monopole is
spherically symmetric as it is shown on the left hand side of the
Fig.~\ref{Fig/Pg4/1:sphere}. If we now place the monopole in the
superconducting medium, then the magnetic field would be pushed out and
spherical symmetry would break down as is depicted on the right hand side
of the Fig.~\ref{Fig/Pg4/1:sphere}.  The region where the magnetic lines
cross a sphere surrounding the monopole will be nothing else than a core
of a vortex on this sphere. The magnetic flux through the sphere is $4\pi
g_M$. From the other hand, the magnetic flux calculated on the solutions
(\ref{Eq/Pg2/1:sphere}) and (\ref{Eq/Pg3/1A:sphere}) is $2\pi /e$.
Comparing these two fluxes one finds $e=1/(2g_M)$, that is $e$ is the
minimal quantum of charge in a full agreement with
Eq.~(\ref{Eq/Pg4/1:sphere}).

Let us note that spontaneous compactification on a monopole
configuration is well-known in Kaluza-Klein models (see, for instance,
Refs.~\cite{monop}). There, fermionic zero modes appear which are not
localized at any particular point of the sphere. Our approach with a
vortex allows one to use a similar mechanism in LED models and to
study the hierarchy of the fermion masses.

\EPSFIGURE[ht]{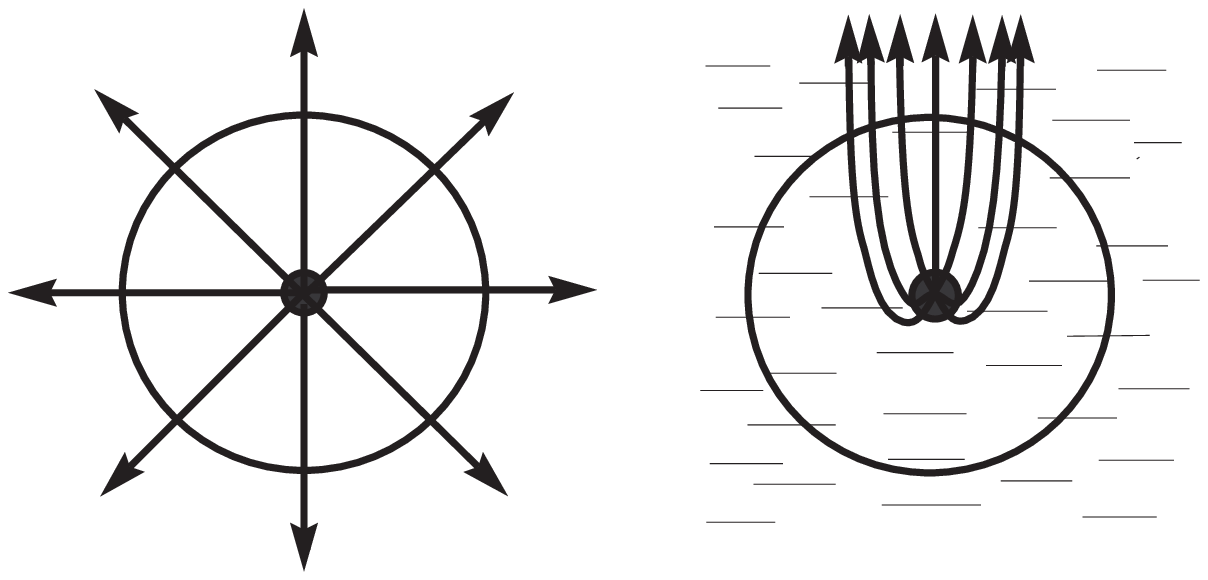,width=100mm}{A monopole in the vacuum (left)
and in the superconducting medium (right).
\label{Fig/Pg4/1:sphere}
}

As we have seen, to describe the vortex configuration on $S^2$ it is
necessary to introduce two patches on the sphere. In practical
calculations it is more convenient, however, to deal with the singular
ansatz (\ref{Eq/Pg2/1:sphere}). To achieve this, one may send the size of
the Southern patch to zero and essentially introduce the Dirac string. In
that case one obtains from Eqs.~(\ref{Eq/Pg3/2:sphere}),
(\ref{Eq/Pg3/3:sphere}), (\ref{Eq/Pg3/5:sphere}) the following boundary
conditions:
\begin{equation}
A(\pi )=1\,,\ \ \ \ \partial _\theta F(\pi )=0.
\label{Eq/Pg4/2:sphere}
\end{equation}
Solving the equations (\ref{Eqn/Pg3/1A:sphere}), (\ref{Eqn/Pg3/2:sphere})
with these boundary conditions one finds the behaviour of $A$ and $F$ near
the North and the South poles ($\xi =\pi -\theta $):
\begin{equation}
A(\theta )=C_{0A}\theta ^2\,, \ \ \ \ F(\theta )=C_{0F}\theta \ \
\mbox{at}\ \ \theta \to 0;
\label{Eq/Pg5/1:sphere}
\end{equation}
\begin{equation}
A(\xi  )=1-C_{\pi A}\xi  ^2\,, \ \ \ \ F(\xi  )=C_{\pi F}\left(1-\lambda
\frac{v^2-C_{\pi F}^2}{4} \xi ^2\right) \ \ \mbox{at}\ \ \theta \to \pi,
\label{Eq/Pg5/2:sphere}
\end{equation}
where $C_{0A}$, $C_{0F}$, $C_{\pi A}$, and $C_{\pi F}$ are unknown
constants ($C_{\pi F}<v$). Note that the number of the unknown constants
is enough to match the  solutions (\ref{Eq/Pg5/1:sphere}),
(\ref{Eq/Pg5/2:sphere}) at some interior point. This  in turn is a strong
argument that the solution to Eqs.~(\ref{Eqn/Pg3/1A:sphere}),
(\ref{Eqn/Pg3/2:sphere}) with boundary conditions (\ref{Eq/Pg4/2:sphere})
does exist.

To complete this section we note that, in the spirit of the Witten's
work~\cite{Witten:1985eb}, a scalar field can be introduced which has a
non-zero value near the North pole and vanishes at the South pole. This
Higgs field is required in the LED models~\cite{Libanov:2000uf} to give
masses to fermionic zero modes. The lagrangian of this field is
\[
{\cal L}_H= \sqrt{-\det(g_{AB})}\left((D^{A}H) ^\dag D_{A}H
-\frac{\kappa  }{2}(|H |^2-\mu ^2)^2-h^2|H|^2|\Phi |^2\right),
\]
where $D_A H =(\partial _A+ieA_A)H $. Making use of the arguments of
Ref.~\cite{Witten:1985eb}, one can show that with $h^2v^2>\kappa \mu ^2$,
there is a stable solution to the field equations $H(\theta )$ which
depends only on $\theta $. This solution is non-zero at the North pole,
$0<H(0)<\mu $, and is zero at the South pole, $H(\xi )=C_{\pi H}\xi ^2$ at
$\theta \to \pi $. The typical size of the solution is of order of the
typical size of the field $\Phi $ (for details, see
Ref.~\cite{Libanov:2002tm} where the flat case is discussed).

\section{Fermionic zero modes in the vortex background.}
\label{Section/Pg5/1:sphere/Fermionic zero modes in the vortex
background.}
In this section we consider fermions and their zero modes in
the vortex background on $S^2$. As in the previous section, we work on
$M^4\times S^2$ manifold. Fermions on this manifold may be represented by
eight-component spinors. We use the chiral representation for the
six-dimensional flat Dirac $\Gamma $-matrices (see
Ref.~\cite{Libanov:2000uf} for notations). In particular, $\Gamma _7={\rm
diag}({\bf 1},{\bf -1})$ is a six-dimensional analog of the
four-dimensional $\gamma _5$.

In the case of the flat space, to obtain fermionic zero modes one usually
considers fermions with a half-integer axial charge with respect to the
vortex group:
\begin{equation}
\Psi '={\rm e}^{i\frac{\Gamma _7}{2}\alpha }\Psi.
\label{Eq/Pg6/1:sphere}
\end{equation}
However, as we have seen in the previous section, in the case of the
vortex on a sphere it is inconsistent to introduce fields with the
half-integer charge. To avoid this difficulty, we change the
representation (\ref{Eq/Pg6/1:sphere}) and consider fermions transforming
as
\begin{equation}
\Psi '={\rm e}^{i\frac{(1+\Gamma _7) }{2}k\alpha }\Psi,
\label{Eq/Pg6/2:sphere}
\end{equation}
where $k$ is a positive integer number. Then, both Weyl spinors which
comprise the Dirac spinor $\Psi$, have integer charges in agreement with
Eq.~(\ref{Eq/Pg4/1:sphere}). In what follows, we will demonstrate that
there are exactly $k$ chiral (from the four-dimensional point of view)
fermionic zero modes.

To describe fermions on $M^4\times S^2$ manifold with the metric
(\ref{metric}) we introduce the following {\em sechsbein},
\begin{equation}
h^a_A=(\delta ^a_\mu ,\, \delta ^a_4 R,\, \delta ^a_5R\sin\theta ),
\label{Eq/Pg6/3:sphere}
\end{equation}
where lower case Latin indices $a,b=0,\ldots ,5$ correspond to the flat
tangent six-dimensional Minkowski space. Then the covariant (with respect
to general coordinate transformations) derivative,
\[
\nabla_A\Psi =\left(\partial _A+\frac{1}{4}R_A^{ab}\Gamma _b\Gamma _a
\right),
\]
where the spin connection is defined as
\[
R_A^{ab}=(\nabla _Ah^{Ba})h^{b}_B,
\]
has only one non-trivial component,
\[
\nabla _\varphi =\partial _\varphi +\frac{\cos\theta }{2}\Gamma
^4\Gamma ^5.
\]
All other components coincide with the usual derivatives.

The lagrangian of the fermions which is invariant under both gauge
(\ref{Eq/Pg6/2:sphere}) and general coordinate transformations may be
chosen as
\begin{equation}
{\cal L}_\Psi =\sqrt{-\det(g_{AB})}\left\{i\bar{\Psi}
h^A_a\Gamma^a \!\left(\!\nabla
_A-iek \frac{1+\Gamma _7}{2}A_A\!\right)\!\Psi -g\Phi ^k\bar{\Psi
}\frac{1-\Gamma _7}{2}\Psi-g\Phi ^{*k}\bar{\Psi }\frac{1+\Gamma _7}{2}\Psi
\right\}.
\label{Eq/Pg6/4:sphere}
\end{equation}

We apply now the standard decomposition procedure. Since the vortex
background does not depend on $x_\mu $,  one can separate variables
related to $M^4$ and $S^2$. To this end let us introduce the transverse
Dirac operator in the background (\ref{Eq/Pg2/1:sphere}),
\[
D=\frac{\Gamma ^0}{R}\left\{i\Gamma ^4\partial _\theta +i\frac{\Gamma
^5}{\sin\theta } \left(\partial _\varphi+\frac{\cos\theta }{2}\Gamma
^4\Gamma ^5-ik\frac{1+\Gamma _7}{2}A\right ) -gR\Phi ^k\frac{1-\Gamma
_7}{2}-gR\Phi ^{*k}\frac{1+\Gamma _7}{2}\right\},
\]
and expand any spinor in a set of the eigenvectors $\Psi _m$ of this
operator
\[
D\Psi _m=m\Psi _m
\]
There may exist a set of discrete eigenvalues $m$ with the separation of
order $gv^k$, and the continuous spectrum starting from $m\gtrsim gv^k$.
All these eigenvalues play a role of the mass of the corresponding
four-dimensional excitations (see Ref.~\cite{Libanov:2000uf} for details).
In what follows we assume that the energy scales probed by a
four-dimensional observer are smaller than $gv^k$, and thus even the first
non-zero level is not excited. So, we are interested only in the zero
modes of $D$:
\begin{equation}
D\Psi =0.
\label{Eq/Pg7/3:sphere}
\end{equation}

To solve the equation (\ref{Eq/Pg7/3:sphere}), we first of all separate
$\phi $ and $\theta $-dependence. The operator of the generalized angular
momentum (see, for instance, Ref.~\cite{Jackiw:1975fn}) commuting with $D$
is, in our case,
\[
J=-i\partial _\varphi -k\frac{1+\Gamma _7}{2}.
\]
The eigenvectors of this operator with the antiperiodic boundary
conditions\footnote{The antiperiodic conditions are chosen due to the
sechsbein (\ref{Eq/Pg6/3:sphere}). In this sechsbein the shift on $2\pi $
also assumes the rotation of the sechsbein on $2\pi $ which changes in
turn the sign of fermions.
} are
\begin{equation}
\Psi _n(\theta ,\varphi )=\exp\left[ i\varphi \left(k\frac{1+\Gamma_7
}{2}-n+\frac{1}{2} \right)\right ]\Psi (\theta ,n)= \left(
\begin{array}{l}
{\bf 1}\cdot{\rm e}^{i\varphi \frac{2k-2n+1}{2}} f_1(\theta )\\
{\bf 1}\cdot{\rm e}^{i\varphi \frac{2k-2n+1}{2}} f_2(\theta )\\
{\bf 1}\cdot{\rm e}^{-i\varphi \frac{2n-1}{2}} f_3(\theta )\\
{\bf 1}\cdot{\rm e}^{-i\varphi \frac{2n-1}{2}} f_4(\theta )\\
\end{array}
\right),
\label{Eq/Pg7/5:sphere}
\end{equation}
where $n$ is an integer number and ${\bf 1}=(1,1)^T$ is a two component
spinor.

Substituting Eq.~(\ref{Eq/Pg7/5:sphere}) into Eq.~(\ref{Eq/Pg7/3:sphere}),
one obtains the following two sets of the differential equations,
\begin{equation}
\left\{
\begin{array}{l}
\left[\partial _\theta +\displaystyle\frac{\cot\theta
}{2}-\frac{1}{\sin\theta }\left(n-\frac{1}{2}-k(1-A)\right)
\right]f_1-Qf_4=0,\\
\\
\left[\partial _\theta +\displaystyle\frac{\cot\theta
}{2}+\frac{1}{\sin\theta }\left(n-\frac{1}{2}\right) \right]f_4-Qf_1=0;
\end{array}
\right.
\label{Eq/Pg8/1:sphere}
\end{equation}
\begin{equation}
\left\{
\begin{array}{l}
\left[\partial _\theta +\displaystyle\frac{\cot\theta
}{2}+\frac{1}{\sin\theta }\left(n-\frac{1}{2}-k(1-A)\right)
\right]f_2+Qf_3=0,\\
\\
\left[\partial _\theta +\displaystyle\frac{\cot\theta
}{2}-\frac{1}{\sin\theta }\left(n-\frac{1}{2}\right) \right]f_3+Qf_2=0,
\end{array}
\right.
\label{Eq/Pg8/2:sphere}
\end{equation}
where
\begin{equation}
Q(\theta )=gRF^{k}(\theta ).
\label{Eq/Pg8/3:sphere}
\end{equation}
Note that the equations (\ref{Eq/Pg8/1:sphere}) can be obtained from the
equations (\ref{Eq/Pg8/2:sphere}) by the replacement $n-1\to -n$, $k\to
-k$, $Q\to -Q$, $f_2\to f_1$, $f_3\to f_4$.

Let us consider the equations (\ref{Eq/Pg8/2:sphere}) for $f_{2,3}$. It is
convenient to introduce new functions $g_{2,3}$ defined by
\begin{equation}
f_{2,3}=g_{2,3}\frac{\sin^{(k-n)}\frac{\theta
}{2}}{\cos^{(k-n+1)}\frac{\theta }{2}}\exp\left(-k\int \limits_{}^{\theta
}\frac{A(x)}{\sin x}dx \right).
\label{Eq/Pg9/1:sphere}
\end{equation}
Then the equations for $g_{2,3}$ take the following form (the prime
denotes a derivative with respect to $\theta $):
\begin{eqnarray}
&&g''_2-\left[\frac{Q'}{Q}+\frac{1}{\sin\theta
}(2n-1-k(1-A)) \right]g'_2 -Q^2 g_2=0,
\label{Eqn/Pg9/1A:sphere}\\
&&g_3=-\frac{g'_2}{Q}.
\label{Eqn/Pg9/2:sphere}
\end{eqnarray}

To proceed further, we will need the following theorem.
\begin{theorem}Let $g_2$ satisfy the Eq.
(\ref{Eqn/Pg9/1A:sphere}) with $Q$ which is differentiable and has no
nodes  at the interior points of the interval $0<\theta <\pi $ and $A$
which has no singularities. Assume that at some interior point $\theta
_0$, $g_2$ and $g_2'$ have the same sign. Then $g_2$ and $g_2'$ have the
same sign at all points $\theta _0<\theta <\pi $.
\end{theorem}

\Proof To prove the theorem, one notes that if a function and its
derivative at some point have the positive (negative) sign then at this
point the function grows (falls). This means that the (continuous)
function cannot change its sign before its derivative does it. Let us
assume that there is a point $\pi >\theta_ m>\theta _0$ in which the
derivative changes its sign. This point is nothing else than a local
maximum (minimum) of the function. Since we have supposed that $Q'/Q$ and
$A$ are regular (which is true for the vortex background), the coefficient
before the second term in (\ref{Eqn/Pg9/1A:sphere}) is non-singular at
$\theta _m$ and hence this term  vanishes at $\theta _m$. Due to
Eq.~(\ref{Eqn/Pg9/1A:sphere}), this means that the second derivative is
strictly positive (negative) at $\theta _m$ which is in contradiction with
the statement that this point is a maximum (minimum). Thus we have shown
the validity of Theorem 1.

It follows trivially from the theorem that if $g_2$ and its derivative
have different signs at some point $\theta _0$, then they have different
signs at all points $0<\theta< \theta _0$.

\vspace{8mm} Now we are ready to demonstrate that the equations
(\ref{Eq/Pg8/2:sphere}) possess exactly $k$ normalizable solutions (at
$k>0$) while Eqs.~(\ref{Eq/Pg8/1:sphere}) have no normalizable solution.
Consider first the case $n>0$.

(i) Let us find the behaviour of $g_2$ in the vicinity of the South pole
($\xi =\pi -\theta $, $\xi \to 0$). Near the South pole $Q'\to 0$, $A\to
1$. So, Eq. (\ref{Eqn/Pg9/1A:sphere}) has two linearly independent
solutions,
\begin{equation}
g_{2\pi }^{(1)}=a_\pi \left(1+\frac{Q_\pi }{4n}\xi ^2+\ldots    \right)
\label{Eq/Pg10/1:sphere}
\end{equation}
and
\[
g_{2\pi }^{(2)}=b_\pi \xi ^{2-2n}(1+\ldots ),
\]
where $a_\pi $ and $b_\pi $ are unknown constants, $Q_\pi\equiv Q(\pi)$.
(For $n=1$, $g_{2\pi}^{(2)}=b_\pi \ln \xi $.) The function $g_{2\pi
}^{(2)}$ yields the following behaviour of $f_{3}$,
\[
f_{3\pi }^{(2)}\sim \frac{b_\pi }{\xi ^n},
\]
which is not normalizable at $n>0$. We conclude that $b_\pi =0$.  Thus,
\[
f_{2\pi }\sim a_\pi \xi ^{n-1}\,,\ \ \ f_{3\pi }\sim a_\pi \xi ^n.
\]

(ii) Since only one solution $g_{2\pi }^{(1)}$,
Eq.~(\ref{Eq/Pg10/1:sphere}), survives at the South pole, we are able to
determine the relative sign of the function $g_2$ and its derivative.
Indeed, it follows from Eq. (\ref{Eq/Pg10/1:sphere}) that $g_2$ and $g_2'$
have different signs near the South pole. Due to Theorem 1 this means that
$g_2$ and $g_2'$ have different signs in the whole  interval $0<\theta
<\pi $.

(iii) Let us investigate now the behaviour of $g_2$ in the vicinity of the
North pole ($\theta \to 0$). There (see Eqs. (\ref{Eq/Pg8/3:sphere}),
(\ref{Eq/Pg5/1:sphere})), $Q=C_Q\theta ^k$ and $A=C_{0A}\theta ^2$, where
 $C_Q=gRC_{0F}^k$. Two linearly independent solutions of
Eq.(\ref{Eqn/Pg9/1A:sphere}) are
\begin{equation}
g_{20}^{(1)}=a_0\left(1+\frac{C_Q^2}{4(k+1)(k+1-n)}\theta ^{2k+2}+\ldots
\right)
\label{Eq/Pg10/5:sphere}
\end{equation}
and
\[
g_{20}^{(2)}=b_0\theta ^{2n}\left(1+{\cal O}(\theta ^2) \right),
\]
where $a_0$ and $b_0$ are some unknown constants.

Note that for $n>0$, $g_{20}^{(2)}$ and its derivative have the same sign
at $\theta \to 0$. So, we conclude that in order to satisfy the (ii)
condition, $a_0$ could not be zero. From the other hand, it  follows from
Eqs. (\ref{Eq/Pg10/5:sphere}) and (\ref{Eq/Pg9/1:sphere}) that
\[
f_{20}^{(1)}\sim a_0 \theta ^{k-n}.
\]
Therefore, $f_2$ can be normalized iff $k\geq n$. If the latter
condition holds, then $g_{20}^{(1)}$ and its derivative have the same
sign. Again, to satisfy the (ii) condition, $b_0$ should be non-zero.
Thus, at $\theta \to 0$ the normalizable solution of
(\ref{Eqn/Pg9/1A:sphere}) is
\begin{equation}
g_{20}=g_{20}^{(1)}+g_{20}^{(2)}
\label{Eq/Pg11/1A:sphere}
\end{equation}
with non-zero $a_0$ and $b_0$. Using Eqs.~(\ref{Eq/Pg11/1A:sphere}),
(\ref{Eqn/Pg9/2:sphere}), and (\ref{Eq/Pg9/1:sphere}) we obtaine the
following behaviour of $f_{2,3}$ at small $\theta$:
\begin{equation}
f_2\sim a_0\theta ^{k-n},
\label{F20:sphere}
\end{equation}
\begin{equation}
f_3\sim b_0\theta ^{n -1}
\label{F30:sphere}
\end{equation}
for $0<n\leq k$ which corresponds to exactly $k$ normalizable modes.

The case $n\leq 0$ as well as Eq.~(\ref{Eq/Pg8/1:sphere}) can be
considered in a similar way. Namely, one should find the behaviour of
$g_2$ ($g_1$) at the South pole and make sure that only one solution
survives, then check that this solution and its derivative have different
signs near the South pole. Theorem 1 then implies that the solution and
its derivative have different signs at all points. Then one finds the
behaviour of the solution in the vicinity of the North pole, makes sure
that both linearly independent solutions should contribute in order to
satisfy the "different sign condition", and checks that there is no $n$ at
which the solution is not singular at the North pole.

To conclude this section let us give four notes in turn. First, we have
obtained exactly $k$ modes with non-vanishing $f_2$ and $f_3$ and
vanishing $f_1$ and $f_4$. These modes are all left handed from the
four-dimensional point of view (for the discussion, see
Ref.~\cite{Libanov:2000uf}). The theory with the Lagrangian
(\ref{Eq/Pg6/4:sphere}) in the presence of other gauge fields is anomalous
however (see also Ref.~\cite{Randjbar-Daemi:2003qd}  for a discussion).

. In order to obtain right
handed modes and to make the theory anomaly free, one introduces other
spinors which rotate under the (vortex) gauge transformations as
\begin{equation}
\Upsilon '={\rm e}^{i\frac{1-\Gamma _7}{2}k\alpha }\Upsilon.
\label{Eq/Pg11/1:sphere}
\end{equation}
In that case one finds the following zero modes,
\begin{equation}
\Upsilon _n(\theta ,\varphi )= \left(
\begin{array}{l}
{\bf 1}\cdot{\rm e}^{-i\varphi \frac{2n-1}{2}} f_3(\theta, g_\Upsilon  )\\
0\\
0\\
{\bf 1}\cdot{\rm e}^{i\varphi \frac{2k-2n+1}{2}} f_2(\theta, g_\Upsilon
)\\
\end{array}
\right)
\label{Eq/Pg12/1:sphere}
\end{equation}
where the functions $f_2$ and $f_3$ satisfy Eqs. (\ref{Eq/Pg8/2:sphere})
with a new Yukawa coupling $g_\Upsilon $.

Second, one can also obtain left- and right-handed modes by introducing
spinors with the charges  $ke(-1+\Gamma _7)/2$ and $-ke(1+\Gamma _7)/2$
(instead of (\ref{Eq/Pg6/2:sphere}) and  (\ref{Eq/Pg11/1:sphere})),
respectively. The new zero mode can be obtained from
(\ref{Eq/Pg7/5:sphere})   and (\ref{Eq/Pg12/1:sphere}) by means of
$CP$-conjugation; for instance, the new right-handed mode is $\Gamma
_0\Gamma _2\Gamma _4\Upsilon ^*$. For the given Yukawa couplings in
(\ref{Eq/Pg12/3:sphere}) this choice actually leads to a different mass
spectrum which we do not consider here.

Third, to localize $k$ modes we have used the vortex with the winding
number one and somewhat unusual\footnote{This is the lowest dimension
interaction allowed by the gauge symmetry for our choice of charges. Note
that in six dimensions, both this interaction and any other Yukawa
interaction are non-renormalizable.} Yukawa coupling in
Eq.~(\ref{Eq/Pg6/4:sphere}).  However, nothing changes in this
construction if one considers the vortex\footnote{The stability of this
solution with respect to splitting into $k$ simple vortices depends on the
parameters of ${\cal{L}}_V$, Eq.~(\ref{Eq/Pg2/1A:sphere}), and is not
discussed here.} with the winding number $k$ and fermions with charge
$e(1+\Gamma _7)/2$. Then one would have a usual Yukawa term $\Phi
\bar{\Psi }(1-\Gamma _7)\Psi $.

Fourth, we have found the behaviour of the zero modes (\ref{F20:sphere}),
(\ref{F30:sphere}) up to unknown constants $a_0$ and $b_0$. These
constants are important for the LED models if the Standard Model fermionic
mass pattern is to be determined. To find $a_0$ and $b_0$ one should use
the normalization condition which reads as
\begin{equation}
1=2\pi R^2\int \limits_{0}^{\pi  }\!d\theta \sin\theta (f_2^2+f_3^2).
\label{Eq/Pg12/2:sphere}
\end{equation}
In the next section, we will estimate $a_0$ and $b_0$ and demonstrate that
the mass hierarchy is indeed reproduced in this model.

\section{Mass hierarchy.}
\label{Section/Pg12/1:sphere/Mass hierarchy.}
In the previous section, we
have found $k$ chiral fermionic zero modes in the vortex background on a
sphere. In the LED models~\cite{Frere:2000dc} these modes are interpreted
as the Standard Model fermions. In particular, at $k=3$ we obtain three
generations of the fermions. These fermions get their masses and mixings
from the Yukawa interaction,
\begin{equation}
HX\bar{\Psi }\frac{1-\Gamma _7}{2}\Upsilon +\epsilon H\Phi \bar{\Psi
}\frac{1-\Gamma _7}{2}\Upsilon,
\label{Eq/Pg12/3:sphere}
\end{equation}
where $H$ is the Standard Model Higgs boson which has a non-trivial
profile on the sphere (see the end of Section
\ref{Section/Pg2/1:sphere/Vortex on a sphere.}), $X$ is an additional
scalar which also has a non-trivial profile similar to the one of $H$, and
$\epsilon $ is a small constant (see
Refs.~\cite{Frere:2000dc,Libanov:2002tm} for details). In this section, we
demonstrate that the hierarchical mass pattern inherent to the flat-space
construction of Ref.~\cite{Frere:2000dc} is reproduced in the spherical
case. For the sake of example we just discuss the case of the diagonal
masses which appear after the integration over the sphere from the first
term in Eq.~(\ref{Eq/Pg12/3:sphere}) (we assume that $g=g_\Upsilon $ for
simplicity),
\begin{equation}
m_{nn}\sim\int \limits_{0}^{\pi }\!d\theta \sin\theta H(\theta )X(\theta
)f_2^2(n,\theta ).
\label{Eq/Pg13/1:sphere}
\end{equation}
The mixings appear from the second term and can be easily treated in a
similar way.

Let $R\theta _\Phi \sim 1/\sqrt{\lambda } v$ be the typical size of $\Phi
$. Then, as it has been pointed out in the
Section~\ref{Section/Pg2/1:sphere/Vortex on a sphere.}, the typical width
of $H$ is also of order $R\theta _\Phi $. The integral in
Eq.~(\ref{Eq/Pg13/1:sphere})  saturates in the region $0<\theta < \theta
_\Phi $. In this region, we can use the leading behavior of $f_2$
(\ref{F20:sphere}). Substituting (\ref{F20:sphere}) into
(\ref{Eq/Pg13/1:sphere}) and assuming that $H\sim H(0)$, one finds
\begin{equation}
m_{nn}\sim a_0(n)^2\theta _\Phi ^{2(k-n+1)}.
\label{Eq/Pg13/2:sphere}
\end{equation}
So, to find masses one should know the constant $a_0(n)$. The latter in
turn requires the knowledge of the explicit solution of
(\ref{Eq/Pg8/2:sphere}) but unfortunately there is no analytical solution
for the vortex. However, we can estimate $a_0$ by using the following
rough approximation for the background.  Let us approach $F$ and $A$ by:
\begin{equation}
F(\theta )= \left[
\begin{array}{l}
\displaystyle\frac{\theta }{\theta _\Phi }C_{\pi F}\ \ \mbox{at} \ \
0<\theta <\theta _\Phi, \\
C_{\pi F}\ \ \mbox{at} \ \ \theta _\Phi <\theta <\pi
\end{array}
\right.
\label{Eq/Pg13/3:sphere}
\end{equation}
and
\begin{equation}
A(\theta )= \left[
\begin{array}{l}
0\ \ \mbox{at} \ \ 0<\theta <\theta _A, \\
1\ \ \mbox{at} \ \ \theta _A <\theta  <\pi,
\end{array}
\right.
\label{Eq/Pg13/3A:sphere}
\end{equation}
where $R\theta _A\sim 1/(ev)$ is the typical size of the gauge core. The
typical width of the fermionic wave function is of order of $\theta _\Psi
=1/Q_\pi =1/(gRC_{\pi F}^k)\simeq1/(gRv^k)$. In what follows, we will work
in the regime
\begin{equation}
\theta _\Phi <\theta _A <\theta _\Psi \ll 1
\label{Eq/Regime:sphere}
\end{equation}
and assume that $k$ is odd because $k=3$ in the most interesting case.

One can solve now Eqs.~(\ref{Eq/Pg8/2:sphere}) in the background
(\ref{Eq/Pg13/3:sphere}), (\ref{Eq/Pg13/3A:sphere}). The calculations are
straightforward but somewhat tedious. We just sketch them here. In each of
the three regions, $0<\theta<\theta_\Phi$; $\theta_\Phi<\theta<\theta_A$;
and  $\theta_A<\theta<\pi$, the most general solution for $g_2(\theta)$ is
an arbitrary linear combination of two independent solutions. At the
intermediate points $\theta=\theta_\Phi$ and $\theta=\theta_A$, we have to
match both $g_2$ and $g_3=-\partial_\theta g_2/Q$, hence four matching
conditions. One more condition is at $\theta=\pi$ (see
Section~\ref{Section/Pg5/1:sphere/Fermionic zero modes in the vortex
background.}) and selects only one of the two linearly independent
solutions at $\theta_A<\theta<\pi$. These conditions, together with the
definitions (\ref{Eq/Pg9/1:sphere}) and Eqs.~(\ref{F20:sphere}),
(\ref{F30:sphere}), allow to express all unknown constants through, say,
$a_0$. The latter is determined by the normalization condition
(\ref{Eq/Pg12/2:sphere}). Technically, one can safely use the perturbative
expansion in $\theta$ for $\theta\lesssim\theta_A\ll 1$ and assume that
$\ln(\theta _A/\theta_\Phi )\sim\ln(\theta _\Psi / \theta _A)\sim 1$. In
this way, we arrive at
\begin{eqnarray}
&&a_0(n)\sim \frac{1}{R}\left[
\begin{array}{l}
\displaystyle\frac{1}{\theta _A^{k-n+1}}B_n \ \ \mbox{at} \ \ k>2n,\\
\\
\displaystyle\frac{1}{\theta_A ^{\frac{k+1}{2}}}\ \ \mbox{at}\ \
n=\frac{k+1}{2},\\
\\
\displaystyle\frac{1}{\theta _A^{k-n+1}}\ \ \mbox{at}\ \ k<2n-1;
\end{array}
\right.
\label{Eq/Pg14/1:sphere}\\
\nonumber\\
&&b_0(n)\sim \frac{1}{R}\left[
\begin{array}{l}
\displaystyle\frac{Q_\pi }{\theta _A^{n-1}} \ \ \mbox{at} \ \ k>2n,\\
\\
\displaystyle\frac{Q_\pi}{\theta_A ^{\frac{k-1}{2}}}\ \
\mbox{at}\ \ n=\frac{k+1}{2},\\
\\
\displaystyle\frac{Q_\pi \theta_\Phi ^{k-2n+2} }{\theta _A^{k-n+1}}\ \
\mbox{at}\ \ k<2n-1,
\end{array}
\right.
\nonumber
\end{eqnarray}
where
\[
B_n=\left[1+\alpha \frac{\theta _A}{\theta _\Psi } \left(\frac{\theta
_A}{\theta _\Phi } \right)^{k-2n}\right]
\]
and $\alpha $ is a constant of order one.

Substituting now Eq.~(\ref{Eq/Pg14/1:sphere}) into
Eq.~(\ref{Eq/Pg13/2:sphere}) and taking $k=3$, one finds
\[
m_{33}:m_{22}:m_{11}=1:\left(\frac{\theta _\Phi }{\theta _A}
\right)^2:B_1^2\left(\frac{\theta _\Phi }{\theta _A}
\right)^4=1:\left(\frac{e^2 }{\lambda } \right):\left( \frac{e^2 }{\lambda
} \right)^2
\]
where in the last equality we assume $\theta _A^2/(\theta _\Psi \theta
_\Phi )=(\sqrt{\lambda }gv^{2})/e^2<1$ which does not contradict to
(\ref{Eq/Regime:sphere}).

This hierarchy is somewhat different from the "flat"
case~\cite{Libanov:2002tm} because in the present case, the hierarchy does
not depend on $g$ and is governed by the small parameter $(\theta _\Phi
/\theta _A)^2$ while in the "flat" case it depends on $g$: a small
parameter is $R_\Phi /R_\Psi $. This difference is a consequence of
different charge assignements.

\section{Conclusion.}

To conclude, we constructed a vortex-like solution on a two dimensional
sphere. We investigated fermionic zero modes in the vortex background and
demonstrated that there are exactly $k$ chiral zero modes, where $k$
is the topological number of the effective background. We have also
demonstrated that this construction can be used in the Large Extra
Dimensions model building. In particular, the hierarchical fermionic mass
pattern is reproduced in this model. The compactification does away with
the need to confine gauge fields (a difficult enterprise by any account) .
The present construction further also avoids spurious fermions. While
compactification on a disk, localising left modes on the vortex introduces
spurious right fermions at the edge: such states could be exploited to
reduce the number of neutrino fields needed~\cite{Frere:2001ug}, but could
prove a disaster for other fermions; ordinary quarks for instance could
annihilate into a pair of such unwanted modes.

We are indebted to S.~Dubovsky and V.~Rubakov for numerous helpful
discussions and to M.~Shaposhnikov who called our attention to this
problem.  This work is supported in part by the IISN (Belgium), the
``Communaut\'e Fran\c{c}aise de Belgique''(ARC), and the Belgium Federal
Government (IUAP). S.T. acknowledges warm hospitality of the Institute for
Nuclear Theory, University of Washington (Seattle), at the final stages of
the work. The work is also supported in part by RFFI grant 02-02-17398
(M.L., E.N.\ and S.T.), by the programme SCOPES of the Swiss NSF, project
No.~7SUPJ062239, financed by Federal Department of Foreign affairs (M.L.\
and S.T.), by INTAS grant YSF 2001/2-129 (S.T.) and by a fellowship of the
``Dynasty'' foundation (awarded by the Scientific Council  of ICFPM)
(S.T.).


\begin{thebibliography}{99}
\bibitem{Jackiw:1981ee}
R.~Jackiw and P.~Rossi,
\npb{190}{1981}{681}.


\bibitem{RuSha}
K.~Akama,
\newjournal{Lect.\ Notes Phys.}{LNPHA}{176}{1982}{267}
[\hepth{0001113}];
V.~A.~Rubakov and M.~E.~Shaposhnikov,
\plb{125}{1983}{136};
G.~W.~Gibbons and D.~L.~Wiltshire,
\npb{287}{1987}{717};
I.~Antoniadis,
\plb{246}{1990}{377};
A.~Nakamura and K.~Shiraishi,
\appol{B21}{1990}{11}.

\bibitem{Libanov:2000uf}
M.~V.~Libanov and S.~V.~Troitsky,
\npb{599}{2001}{319}
[\hepph{0011095}].

\bibitem{Frere:2000dc}
J.~M.~Frere, M.~V.~Libanov and S.~V.~Troitsky,
\plb{512}{2001}{169}
[\hepph{0012306}].

\bibitem{Frere:2001ug}
J.~M.~Frere, M.~V.~Libanov and S.~V.~Troitsky,
\jhep{0111}{2001}{025}
[\hepph{0110045}].

\bibitem{Libanov:2002tm}
M.~V.~Libanov and E.~Y.~Nougaev,
\jhep{0204}{2002}{055}
[\hepph{0201162}].

\bibitem{Gravity}
M.~Giovannini, H.~Meyer and M.~E.~Shaposhnikov,
\npb{619}{2001}{615}
\hepth{0104118};
M.~Giovannini, J.~V.~Le Be and S.~Riederer,
\cqg{19}{2002}{3357}
\hepth{0205222};
M.~Giovannini,
\prd{66}{2002}{044016}
\hepth{0205139}.


\bibitem{Randjbar-Daemi:2003qd}
S.~Randjbar-Daemi and M.~Shaposhnikov,
\jhep{0304}{2003}{016}
[\hepth{0303247}].


\bibitem{Wu:es}
T.~T.~Wu and C.~N.~Yang,
\prd{12}{1975}{3845}.

\bibitem{monop}
Z.~Horvath {\it et.al.},
\npb{127}{1977}{57};
S.~Randjbar-Daemi, A.~Salam and J.~Strathdee,
\npb{214}{1983}{491}.

\bibitem{Witten:1985eb}
E.~Witten,
\npb{249}{1985}{557}.

\bibitem{Jackiw:1975fn}
R.~Jackiw and C.~Rebbi,
\prd{13}{1976}{3398}.
\end{thebibliography}
\end{document}